# Optical Phase Dropout in Diffractive Deep Neural Network

Yong-Liang Xiao

School of Physics and Optoelectronic Engineering, Xiangtan University, Xiangtan 411105, China

ylxiao@xtu.edu.cn

**Abstract:** Unitary learning is a backpropagation that serves to unitary weights update in deep complex-valued neural network with full connections, meeting a physical unitary prior in diffractive deep neural network (**[DN]²**). However, the square matrix property of unitary weights induces that the function signal has a limited dimension that could not generalize well. To address the overfitting problem that comes from the small samples loaded to **[DN]²**, an optical phase dropout trick is implemented. Phase dropout in unitary space that is evolved from a complex dropout and has a statistical inference is formulated for the first time. A synthetic mask recreated from random point apertures with random phase-shifting and its smothered modulation tailors the redundant links through incompletely sampling the input optical field at each diffractive layer. The physical features about the synthetic mask using different nonlinear activations are elucidated in detail. The equivalence between digital and diffractive model determines compound modulations that could successfully circumvent the nonlinear activations physically implemented in **[DN]²**. The numerical experiments verify the superiority of optical phase dropout in **[DN]²** to enhance accuracy in 2D classification and recognition tasks-oriented.

**Key words:** Unitary learning; Diffraction;Phase dropout; Modulation;

Diffractive deep neural network (**[DN]²**) has achieved remarkable development nowadays benefiting on the fact that matrix multiplication could be optically executed in parallel at speed of light with low power consumption as well as high bandwidth. The property of linear superposition in coherent diffraction fulfills a fundamental demand in a multilayer perceptron [1-4]. Thus, coherent diffraction, as a matter of optical mechanism, could provide an elegant manner for fully connecting complex-valued neurons [5].

Recently, instead of phase retrieval in a cascade diffractive architecture [6], backpropagation that manifests a statistical inference [7-9] has been introduced into coherent diffraction for intelligent learning. But, the learning with a separate-type double-channel manner using real-valued backpropagation neglects the intrinsic physical attributes, since the whole optical field involving amplitude and phase is always propagated in free space as a physical entity in diffraction. Thus, complex-valued backpropagation in a single channel is judicious and desirable. Regretfully, there has been a prolonged dilemma on the format of nonlinear activations in complex-valued backpropagation for over three decades, which derives from the differentiability of nonlinear activations in complex space [10]. Plusing ubiquitous unitary property of coherent diffraction in free space, unitary learning is a valid avenue for **[DN]²**. Unitary learning, covering its real counterpart, provides a single channel formulation for **[DN]²** with unitary constraints [11] on complex-valued weights. The adjoint compatible condition ensures that the fundamental sigmoid, tanh, quasi-Relu in complex space could be available as nonlinear activations through conjugation substitution significance (CSS).

However, the square matrix property of unitary weights induces that the function signal has a limited dimension that could not generalize well, in a sense that training is performed on small samples, it typically will operate poorly on the tested data, which implies that the prediction on training data nearly is worse than on the tested data. Such an overfitting drawback could be addressed by a trick called dropout that randomly tailors the links at input space in each layer [12]. We implement an optical phase dropout in unitary learning to tackle an overfitting problem on small samples training in **[DN]²**. The learned random point apertures under Bernoulli distribution with a certain probability is introduced to smother a computational modulation mask for recreating a synthetic mask that randomly selects the conjured connections in each layer. The passing pinholes of random point apertures under Bernoulli distribution are the adjoint parameters in the unitary dropout, and its passing metaphor would be determined once the training is terminated. The phase-only of the random passing pinholes is also randomly generated. The double-random and double-sparse characteristics on the synthetic mask are revealed.

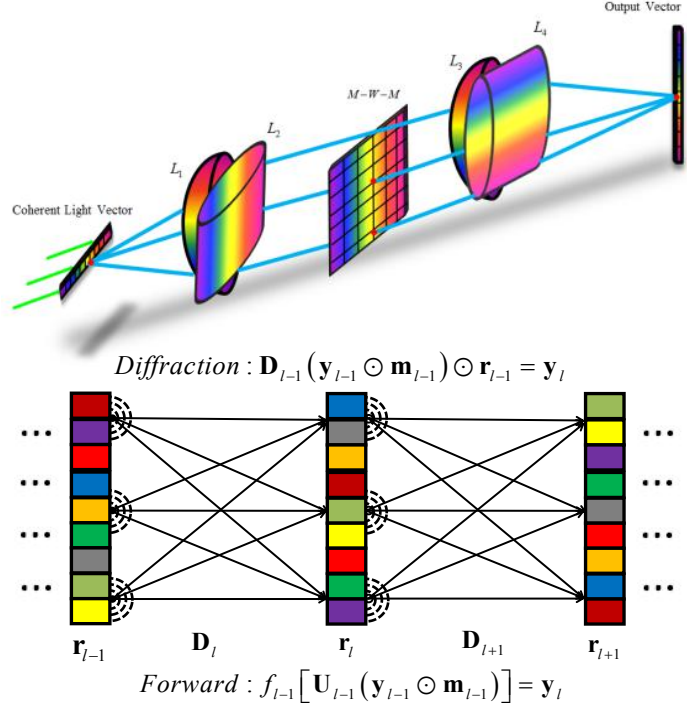

**Fig.1** Schematic of optical intelligent unit in diffractive deep neural network and its analogous portrait

[DN]² is build based on an analogous portrait of cascade diffraction architecture. The optical intelligent unit is shown in Figure 1 for a complex-valued multilayer perceptron, in which a row of signal is paralleled illuminated by coherent light. The combination of spherical and cylindrical lens could project a sole pixel of input vector onto a column of $N \times N$ mask, so that the input vector could be diffused all over a weight mask $M-W-M$ that could be a modulation for weight. The matrix-vector multiplication is optically performed under diffraction. Coherent diffraction and digital unitary network in the $nth$ layer can be modulated to be equivalent. $\mathbf{D}_n$ is used to represent a discrete projection operator with respect to a certain diffraction distance [13]. The diffractive and digital unitary network with an optical phase dropout are respectively demonstrated as

*Digital Neural Network Phase Dropout*     *Deep Diffraction Network Phase Dropout*

$$\begin{cases} f_1\left[\mathbf{U}_1\left(\mathbf{y}_0 \odot \mathbf{m}_0\right)\right] = \mathbf{y}_1 \\ \vdots \quad\quad \vdots \\ f_{n-1}\left[\mathbf{U}_{n-1}\left(\mathbf{y}_{n-2} \odot \mathbf{m}_{n-2}\right)\right] = \mathbf{y}_{n-1} \\ f_n\left[\mathbf{U}_n\left(\mathbf{y}_{n-1} \odot \mathbf{m}_{n-1}\right)\right] = \mathbf{y}_n \end{cases} \quad\quad \begin{cases} \mathbf{D}_1\left(\mathbf{y}_0 \odot \mathbf{m}_0\right) \odot \mathbf{r}_1 = \mathbf{y}_1 \\ \vdots \quad\quad \vdots \\ \mathbf{D}_{n-1}\left(\mathbf{y}_{n-2} \odot \mathbf{m}_{n-2}\right) \odot \mathbf{r}_{n-1} = \mathbf{y}_{n-1} \\ \mathbf{D}_n\left(\mathbf{y}_{n-1} \odot \mathbf{m}_{n-1}\right) \odot \mathbf{r}_n = \mathbf{y}_n \end{cases}$$

Where, $\mathbf{U}_i$ is a unitary weight, $\mathbf{m}_0, \mathbf{m}_1, \mathbf{m}_2 \cdots \mathbf{m}_n$ are the learned phase-only sparse dropout under Bernoulli distribution with random phase-shifting, and $\mathbf{r}_1, \mathbf{r}_2 \cdots \mathbf{r}_{n-1}, \mathbf{r}_n$ are the complex-amplitude modulations in physical layers. Thus, the modulation could be obtained in terms of the equivalence in the intermediate $kth$ layer as

$$f_k\left[\mathbf{U}_k\left(\mathbf{y}_{k-1} \odot \mathbf{m}_{k-1}\right) \odot \mathbf{r}_k\right] = \mathbf{D}_k\left(\mathbf{y}_{k-1} \odot \mathbf{m}_{k-1}\right) \odot \mathbf{r}_k \quad (1)$$

The corresponding multilayer optical mechanism is shown in Figure 2, which is a cascade diffractive architecture and could circumvent nonlinear activations by modulation in terms of the equivalences between digital and diffractive phase dropout. The synthetic mask, implicitly relating to nonlinear activation, is designed as a compound modulation in each layer, and involves a computational modulation as well as a phase dropout. Its output end in $kth$ layer is represented as $\mathbf{y}_k$, which is regarded as an output of optical field passing through a modulation $\mathbf{r}_k$, as well as the input optical field entering into phase dropout layer $\mathbf{m}_k$. The synthetic mask is reconfigurable with a prior unitary phase dropout learned from small samples dataset.

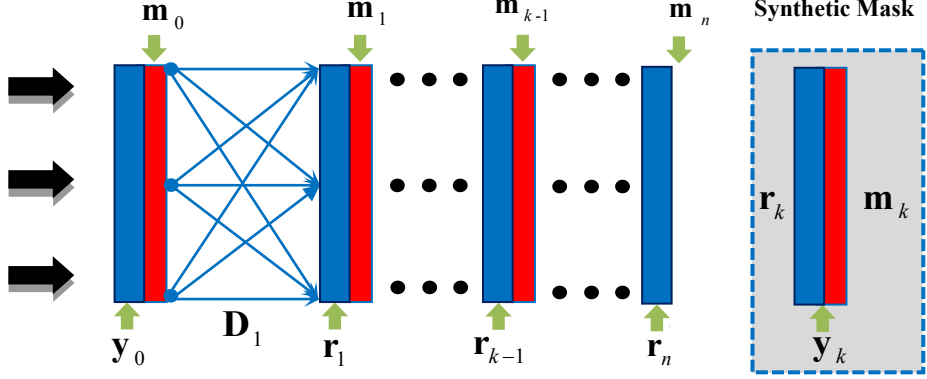

**Fig.2** Multilayer mechanism for diffractive deep neural network and its synthetic mask

The sum of energy of errors between output and target optical field involving amplitude and phase is formulated as a loss function. We can compute the complex variables derivative of loss function with respect to unitary weights using complex-valued chain rules, with the help of unveiled compatible condition for selecting nonlinear activations $f_k^*(\mathbf{z}) = f_k(\mathbf{z}^*)$ [14]. The backpropagation and update in unitary space with phase dropout, are demonstrated as follows

$$\nabla_{\mathbf{w}_k^*}\varepsilon[\varepsilon^*] = -\varepsilon \nabla_{\mathbf{w}_k^*} f(\mathbf{net}_{l+1}^*) \frac{\partial[\mathbf{net}_{\ell+1}^*]}{\partial \mathbf{x}_\ell^*} \frac{\partial[\mathbf{x}_\ell^*]}{\partial \mathbf{net}_\ell^*} \frac{\partial[\mathbf{net}_\ell^*]}{\partial \mathbf{x}_{\ell-1}^*} \frac{\partial[\mathbf{x}_{\ell-1}^*]}{\partial \mathbf{net}_{\ell-1}^*} \cdots \frac{\partial[\mathbf{net}_{k+2}^*]}{\partial \mathbf{x}_{k+1}^*} \frac{\partial[\mathbf{x}_{k+1}^*]}{\partial \mathbf{net}_{k+1}^*} \frac{\partial[\mathbf{net}_{k+1}^*]}{\partial \mathbf{x}_k^*} \frac{\partial[\mathbf{x}_k^*]}{\partial \mathbf{net}_k^*} \frac{\partial[\mathbf{net}_k^*]}{\partial \mathbf{W}_k^*}$$

$$compatible\ condition\ f(\mathbf{net}_{l+1}^*) = f^*(\mathbf{net}_{l+1}^*),$$

$$\frac{\partial[\mathbf{net}_{\ell+1}^*]}{\partial \mathbf{x}_\ell^*} = \frac{\partial[\mathbf{w}_{\ell+1}^*(\mathbf{x}_\ell^* \odot \mathbf{m}_\ell^*)]}{\partial \mathbf{x}_\ell^*} = \mathbf{W}_{\ell+1}^H \odot \mathbf{m}_\ell^*, \frac{\partial[\mathbf{net}_k^*]}{\partial \mathbf{W}_k^*} = [\mathbf{x}_{k-1} \odot \mathbf{m}_{k-1}]^H,$$

$$\Delta \mathbf{W}_k = \mu\{\mathbf{W}_{k+1}^H \mathbf{W}_{k+2}^H \cdots \mathbf{W}_\ell^H \mathbf{W}_{\ell+1}^H\}\{[(-\varepsilon)\odot f'(\mathbf{net}_{\ell+1}^*)]\odot f'(\mathbf{net}_\ell^*)\odot \mathbf{m}_\ell^* \odot f'(\mathbf{net}_{\ell-1}^*)\odot \mathbf{m}_{\ell-1}^* \cdots \odot f'(\mathbf{net}_{k+1}^*)\odot \mathbf{m}_{k+1}^* \odot f'(\mathbf{net}_{k+1}^*)\odot \mathbf{m}_k^*\}$$

$$\Delta \mathbf{W}_k = \mu\varepsilon\left[\prod_{t=k+1}^{\ell+1}\mathbf{W}_t^H\right]\left[\Theta_{t=k}^{\ell+1}f'(\mathbf{net}_t^*)\odot \mathbf{m}_t^*\right][\mathbf{x}_{k-1}\odot \mathbf{m}_{k-1}]^H$$

$define\quad \boldsymbol{\delta}_{\ell+1}=(-\varepsilon)\odot f'(\mathbf{net}_{\ell+1}^*)\odot \mathbf{m}_{\ell+1}^*, \boldsymbol{\delta}_\ell = \mathbf{W}_\ell^H\boldsymbol{\delta}_{\ell+1}\odot f'(\mathbf{net}_\ell^*)\odot \mathbf{m}_\ell^*, \boldsymbol{\delta}_{\ell-1} = \mathbf{W}_{\ell-1}^H\boldsymbol{\delta}_\ell\odot f'(\mathbf{net}_{\ell-1}^*)\odot \mathbf{m}_{\ell-1}^*\cdots$

$$\boldsymbol{\delta}_{k+1} = \mathbf{W}_{k+1}^H\boldsymbol{\delta}_{k+2}\odot f'(\mathbf{net}_{k+1}^*)\odot \mathbf{m}_{k+1}^*, \boldsymbol{\delta}_k = \mathbf{W}_k^H\boldsymbol{\delta}_{k+1}\odot f'(\mathbf{net}_k^*)\odot \mathbf{m}_k^*$$

$$\Delta\mathbf{W}_k = -\mu\boldsymbol{\delta}_k\left[(\mathbf{x}_{k-1}\odot \mathbf{m}_{k-1})\right]^H$$

**Complex Dropout**

$\mathbf{x}_\ell = f(\mathbf{z}_\ell), \mathbf{z}_\ell = \mathbf{W}_\ell[\mathbf{x}_{\ell-1}\odot \mathbf{m}_{\ell-1}] + \mathbf{b}_\ell$

$\begin{cases} \boldsymbol{\delta}_L = (\mathbf{z}_L - \mathbf{t}_L)\odot f'(\mathbf{z}_L^*) \\ \boldsymbol{\delta}_\ell = [\mathbf{W}_{\ell+1}^*]^T\boldsymbol{\delta}_{\ell+1}\odot f'(\mathbf{z}_\ell^*)\odot \mathbf{m}_\ell^* \\ \Delta\mathbf{W}_\ell = -\eta\boldsymbol{\delta}_\ell[\mathbf{x}_{\ell-1}^*\odot \mathbf{m}_{\ell-1}^*]^T \\ \Delta\mathbf{b}_\ell = -\eta\boldsymbol{\delta}_\ell \end{cases}$

$\mathbf{x}_\ell, \mathbf{x}_{\ell-1}, z_\ell, z_L, \mathbf{t}_L, \boldsymbol{\delta}_L, \boldsymbol{\delta}_\ell, \mathbf{m}_\ell, \Delta\mathbf{W}_\ell, \Delta\mathbf{b}_\ell, \in \mathbb{Z}$

**Unitary Dropout**

$\mathbf{x}_\ell = f(\mathbf{z}_\ell), \mathbf{z}_\ell = \mathbf{U}_\ell[\mathbf{x}_{\ell-1}\odot \mathbf{m}_{\ell-1}] + \mathbf{b}_\ell, s.t\ \mathbf{U}_\ell\mathbf{U}_\ell^H = \mathbf{U}_\ell^H\mathbf{U}_\ell = \mathbf{I}$

$\begin{cases} \boldsymbol{\delta}_\ell = \mathbf{U}_{\ell+1}^H\boldsymbol{\delta}_{\ell+1}\odot f'(\mathbf{z}_\ell^*)\odot \mathbf{m}_\ell^*, \Delta\mathbf{W}_\ell = -\mu\boldsymbol{\delta}_\ell[\mathbf{x}_{\ell-1}^*\odot \mathbf{m}_{\ell-1}^*]^T \\ \mathbf{G}_\ell = \mathbf{U}_\ell\Delta\mathbf{W}_\ell^H - \Delta\mathbf{W}_\ell\mathbf{U}_\ell^H \\ \Delta\hat{\mathbf{U}}_\ell = \exp[-\lambda\mathbf{G}_\ell] \\ \Delta\mathbf{b}_\ell = -\mu\boldsymbol{\delta}_\ell \end{cases}$

$\mathbf{x}_\ell, \mathbf{x}_{\ell-1}, \mathbf{z}_\ell, \boldsymbol{\delta}_\ell, \mathbf{m}_\ell, \mathbf{U}_\ell, \Delta\mathbf{W}_\ell, \Delta\hat{\mathbf{U}}_\ell, \mathbf{G}_\ell, \Delta\mathbf{b}_\ell, \in \mathbb{Z}$

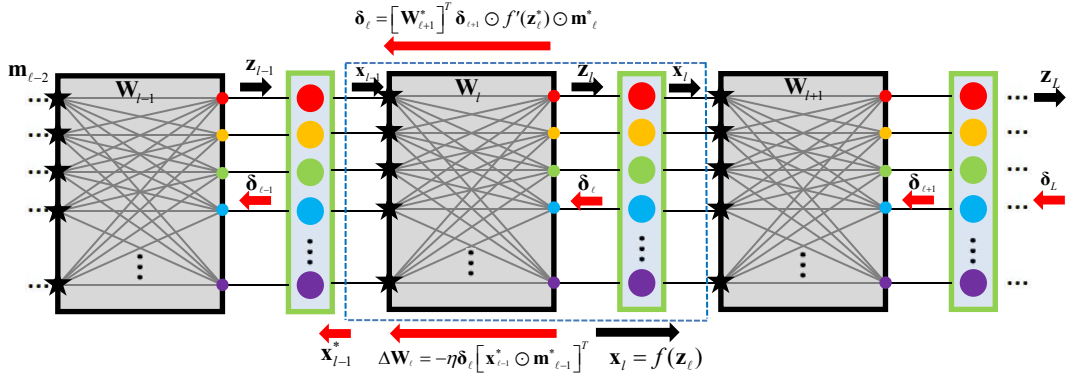

**Fig.3** Schematic diagram of backpropagation in the complex dropout, ★ denotes the phase dropout

Unitary dropout is evolved from the complex dropout, the CSS is implemented in its real backpropagation counterpart. The main ingredient in unitary dropout is the complex dropout, the gradient translated from Euclidean to Riemannian space is fed by complex dropout. The schematic diagram of backpropagation in complex dropout is shown in Figure 3, which is different from a typical complex space by multiplying the phase dropout term $\mathbf{m}_\ell$. The dropout introduced into the input field at each layer induces that the backpropagation errors during epochs should be modulated with the conjugation of phase dropout $\mathbf{m}_\ell^*$, which implies that an optical inverse propagation with a certain random phase-shifting. The phase dropout would present a double-random characteristic with respect to the passing places and its phase-only values that have a sparse trait. As to unitary dropout, shown in Figure 2, $\mathbf{m}_k$ integrated $\mathbf{x}_{\ell-1}$ is regarded as a holistic input optical field in $[DN]^2$, primarily decides the modulation term $\mathbf{r}_{k+1}$. $\mathbf{m}_\ell^*$, just arising in Euclidean gradient, is imparted to the weights update $\Delta \mathbf{W}_\ell$ to generate Riemannian gradient, the evolutional format is similar to unitary learning without phase dropout, which typically presents statistical inference.

In real space, dropout is a blind tailor that randomly trims the redundant connections, enabling the training present variants of parallel computational network. The phase dropout in complex/unitary space could be a tailor but also a random phase shifter that allows for place and value, as well as depends on the filling probability factor. In the practical unitary dropout, the place of zero-pinhole meets the updated Bernoulli distribution with the same probability in each epoch, while the only-phase also meets random distributions and retains unchanged in each epoch once generated inchoatively. We interpret the synthetic mask in Figure 4 with a sandwich structure to exhibit its physical features in $[DN]^2$, in which the modulation and phase dropout are seamlessly compounded.

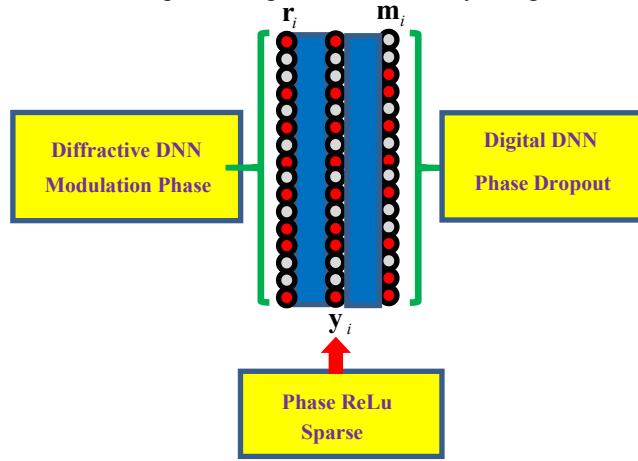

**Fig.4** The schematic diagram of the synthetic mask

In terms of compatible condition, the fundamental sigmoid, tanh and qusi-ReLu in complex space could be adopted in digital unitary dropout. When a standard phase ReLu is utilized as a nonlinear synapse, the modulation is computed as

$$\begin{cases} \mathbf{y}_k = f_k(\mathbf{z} = |\mathbf{A}_z|e^{i\theta_z}) = \begin{cases} \mathbf{z}, \theta_z \in [-\pi/2, \pi/2] \\ 0, otherelse \end{cases}, \mathbf{z} \in \mathbb{Z} \\ \mathbf{r}_k = \frac{\mathbf{y}_k}{\left[\mathbf{D}_k(\mathbf{y}_{k-1} \odot \mathbf{m}_{k-1})\right]} \end{cases} \quad (2)$$

There would be a serial of zero-pinholes arising in the output $\mathbf{y}_k$, to which a sparse metaphor for modulation $\mathbf{r}_k$ is the same. And the distributions of zero-pinhole could be fabricated easily by tuning the hyper-parameter filling probability factor, because $\mathbf{D}_k$ is a diffractive operator albeit the double-sparse input is presented in $kth$ layer. To guarantee the modulation is computational, it is better to choose a big probability factor in the input layer to realize an efficient diffractive diffusion, or else the modulation in the subsequent layer could be fairly massive. So, it is suitable for a natural rather than binary images. If the nonlinear activations adopted in digital unitary learning is sigmoid and tanh, whether there is a sparse trait in modulation $\mathbf{r}_k$ largely depends on the setting probability factor in $\mathbf{m}_{k-1}$ because the diffusion effect of diffraction operator $\mathbf{D}_k$ is weak. Seen from Figure 4, the synthetic mask in Eq.(2) renders an interesting characteristic that the sparse property is enhanced with a standard phase ReLu and its phase dropout. The double-random property for $\mathbf{y}_k$ is reflected in the place of zero-pinhole and the value of only-phase.

We continue to occupy in nonlinear activations implemented in a complex dropout for feeding optical phase unitary dropout. Numerical simulations to do researches on the nonlinear activations selection for a unitary dropout. In terms of compatible condition, sigmoid, tanh and ReLu are available in complex space with CSS, and it is necessary to ascertain the convergent performance. We observe the learning curves in a real 4 × 4 XOR logic [15-16] training in complex/unitary space (**Supplementary materials**). In a common complex phase dropout, Tanh requires a very small dropFraction to guarantee the convergence, a dropoutFraction in Phase ReLu is larger, the convergent speed is faster, furthermore Sigmoid ensures convergent but exhibits a strong uncertainty. In unitary dropout, additionally, Sigmoid and Tanh present a serial of fluctuations during iterations and phaseReLu is still tough. In the words, all the numerical simulations above manifest that phase ReLu with large phase dropout Fraction is a good choice in complex/unitary dropout.

For unitary phase dropout applications, **[DN]²** with a synthetic mask modulation fed by the digital unitary phase dropout is presented, and could have a statistical ubiquility. Angular spectrum propagation [17] is specifically implemented for Fresnel diffraction, in which planar wave with different frequency is an Eigent state of coherent propagation with a tilted angle. We generate a small diffractive samples involving amplitude and phase to feed digital unitary phase dropout and produce a synthetic mask described in Figure 2. The situation that intensity mapping to intensity is also tillable because there are not any compulsory complex-valued constraints on samples.

2D classification and verification are a fundamental task that could be operated with deep neural network and has achieved high accuracy with real-valued backpropagation [18]. **[DN]²** in unitary space could provide high degree of freedom in wavelength, distance, phase for intelligent manipulation and modulation. Here, the CIFAR-10 dataset is implemented in unitary dropout, sigmoid is chosen for scoring whereas phase ReLu is chosen as an activation at hidden layer. We investigate the performance of a two-layer **[DN]²**, five selected images in CIFAR-10 are displayed in Figure 5(a) and the corresponding diffractive field with angular spectrum propagation is generated to feed the training with small samples. The training results of a standard unitary phase dropout are performed to fabricate the compounded synthetic masks that involve modulation and phase dropout in terms of Eq.(2). The learning masks with phase drop out are exhibited in Figure 5(a). The learning rate is set as 0.05, the filling probability fraction is given as 0.9, and the initialization is a common complex-valued respectively containing unitary matrices in real and imaginary parts. The learning is operated on the platform of Python version 3.6.7 using a desktop computer (Intel Core i7-4510U CPU at 2.60 GHz), and the achieved modulation and phase dropout mask are shown in Figure 5(b). For comparisons, the learning and vaditation curves with as well as without phase dropout are shown in Figure 6. It is a trick that provides an avenue of approximately combining exponentially many different neural network architectures efficiently, amounting to training a collection of $2^n$ thinned networks with weights sharing [12]. The diffractive recognition at the output imaging plane should be designed with an area encoding for the category indicators and then vectorized. The dimension of the loaded samples is the same to unitary weight, namely a square matrix.

In the proof-of-concept, the pixels of CIFAR-10 are resized to be 64×64, whose vectorization is the dimension of the square unitary weights. It is an example that exploits small samples in deep unitary dropout. There are 4096 vectors of 4096 rows gathered together for feeding. Obviously, dropout in unitary space is necessary and desirable. Seen from Fig. 6, phase dropout presenting double-random and double sparse characteristics largely enhance the classification performance to achieve an accuracy of 87.21% with a dropout Fraction equal to 0.85, compared to the unitary full-connection accuracy of 81.56%.

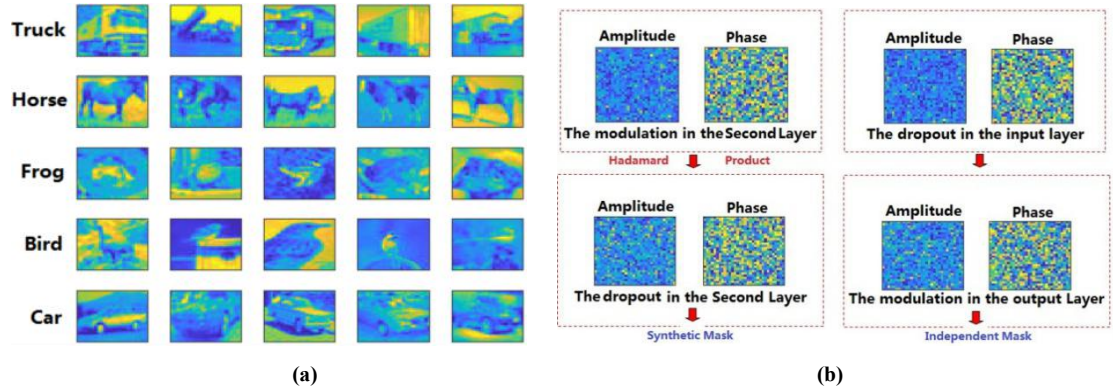

(a)      (b)

**Fig.5** Unitary dropout on the CIFAR-10 for feeding diffractive deep neural network. **(a)** Selected training samples in CIFAR-10, **(b)** The masks computed in a two-layer **[DN]²** are exhibited**,** an independent phase dropout mask is presented at the input layer, a synthetic mask with the phase dropout and modulation is presented at the hidden layer.

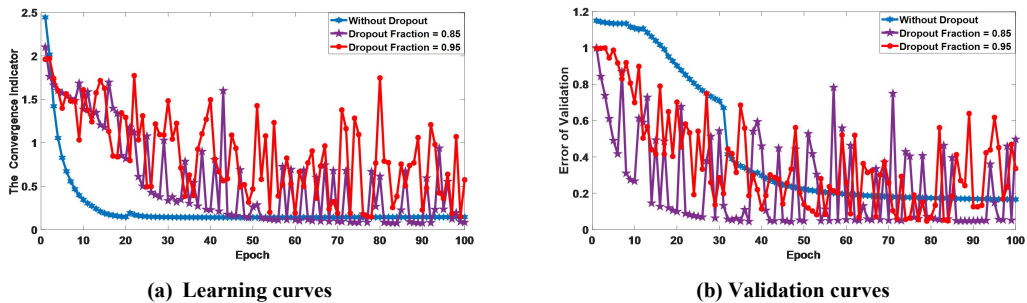

(a) Learning curves      (b) Validation curves

**Fig.6** (a) The learning curves in unitary learning with/without phase dropout, (b) The validation curves in unitary learning with/without phase dropout with respect to (a).

Aiming at the practical training problems in **[DN]²** that small samples are implemented coming from unitary representation in backpropagation with square weights, phase dropout could be realized in optical setup accompanying modulation at output end in each layer, sparse phase coding could implement the learnable dropout apertures with a setting filling probability factor and random phase shifting in optical beam path to operate optical diffraction intelligent predictions, avoiding the overfitting phenomena happened on the tested data that comes from small samples training. The phase dropout trick presents a good generalized ability, more than circumventing nonlinear activations implemented in the potential optical Situ realization. The degenerate format for phase dropout backpropagation is laughed for the first time.

# Supplementary Materials

## The convergent comparisons between complex dropout and unitary dropout

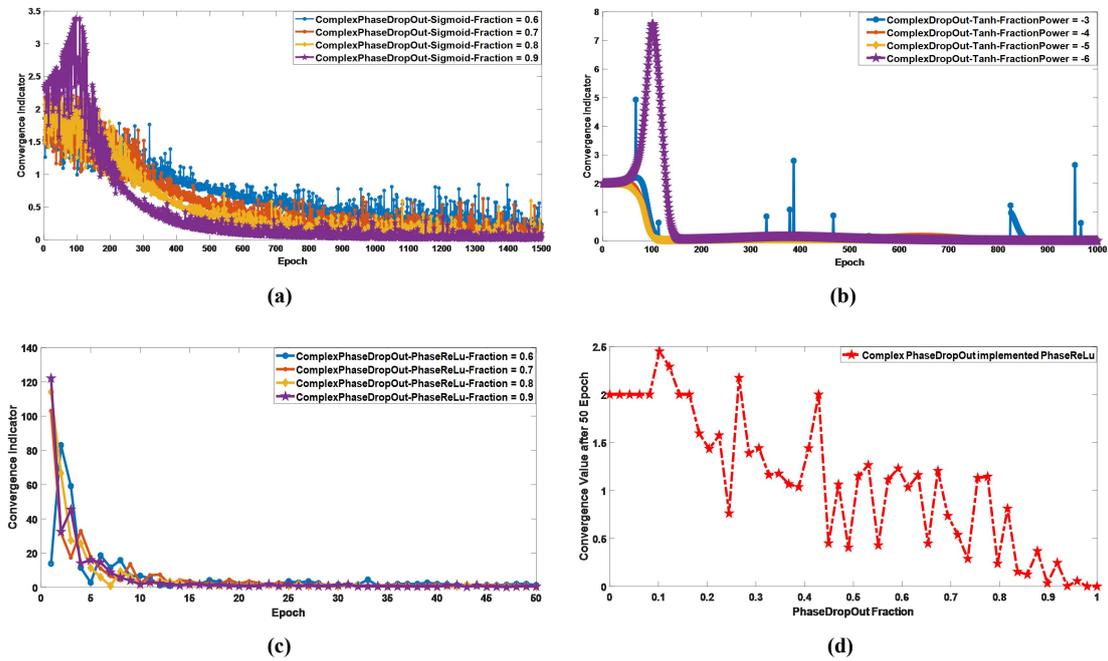

**Fig.1** The learning curves using the fundamental activations in Complex Phase Dropout

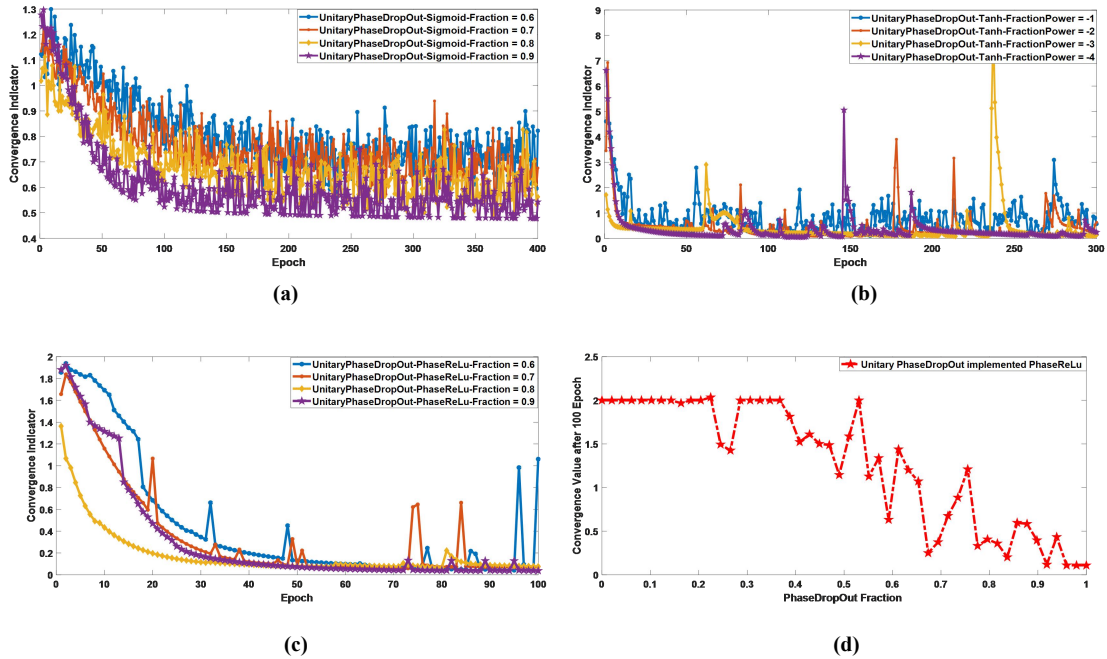

**Fig.2** The learning curves using the fundamental activations in Unitary Phase Dropout

Figure.1(a)(b)(c) are the learning curves that implement fundamental nonlinear activations versus the dropoutFraction in complex dropout. There is a power indicator describing the dropoutFraction as $1-1\times 10P$ for tanh. In Figure.1(d), the curve presents the convergent characteristic with respect to the dropoutFraction, whose value ranges among 0.4~0.5 could lead to convergence. As to unitary phase dropout, Figure.2 (a)(b)(c) display the learning curves. Figure.2(d) exhibits that the larger phase dropout Fraction, implying the weaker sparse trait about the phase dropout, will induce better convergence within fixed epochs.